\documentclass[12pt]{iopart}

\usepackage{iopams}  
\usepackage{graphicx}
\usepackage{hyperref}
\usepackage{color}
\usepackage{epstopdf}
\usepackage{bbold}
\usepackage{array}

\newcommand{\ket}[1]{\ensuremath{|{#1}\rangle}}

\newcolumntype{P}[1]{>{\centering\arraybackslash}p{#1}}

\begin{document}

\title{Neutrino Flavour Evolution Through Fluctuating Matter}

\author{Y. Yang and J. P. Kneller}
\address{Department of Physics, North Carolina State University,Raleigh,North Carolina 27695, USA}

\begin{abstract}
A neutrino propagating through fluctuating matter can experience large amplitude transitions between its states. Such transitions occur in supernovae and compact object mergers due to turbulent matter profiles and neutrino self-interactions. 
In this paper we 
study, both numerically and analytically, three-flavour neutrino transformation through fluctuating matter built from two and three Fourier modes. We find flavor transformation effects which cannot occur with just two flavours. For the case of two Fourier modes we observe the equivalent of ``induced transparency'' from quantum optics whereby transitions between a given pair of states are suppressed due to the presence of a resonant mode between another pair. When we add a third Fourier mode we find a new effect whereby the third mode can manipulate the transition probabilities of the two mode case so as to force complete transparency or, alternatively, restore ``opacity'' meaning the perturbative Hamiltonian regains its ability to induce neutrino flavour transitions. In both applications we find analytic solutions are able to match the amplitude and wavenumber of the numerical results to within a few percent. We then consider a case of turbulence and show how the theory can be used to understand the very different response of a neutrino to what appears to be two, almost identical, instances of turbulence. 
\end{abstract}

\submitto{\jpg}

%
%

\section{Introduction}

Perhaps one of the more unusual cases of a driven quantum mechanical system is the flavour evolution in space/time of a neutrino as it propagates through inhomogeneously distributed matter and/or through a field of other neutrinos. 
The phenomenology of neutrino flavour evolution in environments such as core-collapse supernovae, the merger of two neutron stars or a neutron star and a black hole, has been found to be very rich: for recent reviews of supernova neutrinos see Mirizzi \emph{et al.} \cite{2016NCimR..39....1M} and Horiuchi \& Kneller \cite{2017arXiv170901515H}.
An equivalence of a neutrino to other driven quantum systems can be made because the neutrino flavour evolution in these environments is calculated from a Schr\"{o}dinger equation governed by a Hamiltonian. Due to the difference in the neutrino masses, a $N_f$ flavour neutrino has $N_f$ distinct eigenstates of the Hamiltonian which we can treat just like the eigenstates of an atom or molecule.  Some of the causes of neutrino flavour transformation in supernovae and compact-object merger environments are well understood - e.g. the Mikheyev-Smirnov-Wolfenstein effect \cite{Wolfenstein1977,M&S1986} - however at the present time the flavour transformation induced by turbulence in the matter is only understood for the case of two neutrino flavours, and similarly the phenomenology of neutrino flavour transformation due to `self-interactions' \cite{2006PhRvL..97x1101D} have been solved only in simplified scenarios \cite{2006PhRvD..74j5010H}. In order to understand neutrino flavour evolution in such environments, we need analytical tools which are able to predict the response of a neutrino to such stimulations. 

The effect of matter fluctuations upon neutrinos has received a lot of attention from an analytic perspective. Both Floquet theory and the Rotating Wave Approximation approaches have been used to calculate the effect upon a \emph{two} flavor neutrino of matter fluctuations described by a \emph{single} Fourier mode (FM), and through periodic layers of constant density \cite{Ermilova,Akhmedov,1989PhLB..226..341K,1999NuPhB.538...25A,1998PhLB..434..321P,2001PhRvD..63g3003C,2001PAN....64..787A,Kneller:2012id}. From these studies it has been found that for a given matter structure, at some neutrino energies the probability for the neutrino to transition between its two states can be enhanced via parametric resonance. 
Recently the case of two flavour neutrino evolution through non-constant, non-periodic matter fluctuations, as one would find in a turbulent medium, was considered by Patton, Kneller and McLaughlin (PKM) \cite{2014PhRvD..89g3022P}. 
Note an alternative analysis of the similar problem of a two-level atom interacting with a stochastic electromagnetic field is found in Cummings \cite{1982NCimB..70..102C}. PKM based their theoretical description of the evolution also upon the Rotating Wave Approximation (RWA) and found it gave predictions which were in remarkably good agreement with numerical calculations on a case-by-case basis even though the turbulence is aperiodic. They named their model Stimulated Transitions and found there is a direct correspondence with the predicted response of an irradiated polar molecule \cite{1985PhLA..108..340K,Kondo:1992,Nakai:Meath}. Like the other studies of matter effects upon neutrinos, PKM also saw the effect of parametric resonance but, in addition, they also observed a suppression effect when low frequency / long wavelength modes were present in the turbulence. 

However neutrinos have (at least) three flavours and it is well known that one finds richer phenomenology when a quantum system possesses three or more eigenstates. Perhaps the best known examples are in the field of quantum optics where one observes the phenomenon of electromagnetic induced transparency \cite{1990PhRvL..64.1107H,1991PhRvL..66.2593B,2003RvMP...75..457L,2005RvMP...77..633F} and coherent population trapping into dark states \cite{2012Sci...338.1609D} which has also been seen in quantum dots and solid-state systems \cite{2008NatPh...4..692X,2011Natur.478..497T}. Whether similar effects occur for three flavour neutrinos is presently unknown. 

In our paper we study the effect of matter fluctuations upon a three flavour neutrinos passing through matter fluctuations to a) observe three flavour oscillation phenomena and b) examine the utility of a analytical tool for predicting the response of neutrinos to Hamiltonians which can be decomposed into a Fourier series. In section \S\ref{induced} we study the case of two anharmonic FMs and find the equivalent of electromagnetic induced transparency. 
In section \S\ref{opacity} we add another FM and find a new effect we call Restored Opacity. In both studies we find the analytic solutions and numerical calculations are in excellent agreement. We then finish 
with a case of turbulent matter fluctuations and show how the insight gained from the two and three FM cases can be used to understand why a neutrino can respond so differently to two cases of turbulence which appear, at first glance, to be almost identical. 
Our conclusions and directions for further study are presented in section \S\ref{conclusions}.


\section{Neutrino Propagation Through Fluctuating Matter}
\label{induced}

The problem we wish to solve is the case of a three-flavour neutrino propagating through fluctuating matter. While this is an example of the evolution of a $N$-level quantum system subject to a time-dependent Hamiltonian note that, as commonly found in the literature on neutrino flavour transformation, we switch the variable from time $t$ to position along the neutrino trajectory $r$ where $r=c\,t$ since the neutrino wavepacket is localized in space and typically the energy of neutrinos is much larger than their rest mass hence they move at a speed close to $c$. 
If the neutrino is initially in state $\phi^{(f)}(0)$ in the flavor basis then at position $r$ the neutrino is in the state $\phi^{(f)}(r)$ related to $\phi^{(f)}(0)$ via the evolution matrix $S$ i.e.\ $\phi^{(f)}(r) = S\,\phi^{(f)}(0)$. This matrix can be found by solving the Schr\"{o}dinger equation   
\begin{equation}
\rmi {\frac{\rmd S}{\rmd r}} = H^{(f)}\,S 
\end{equation}
given the initial condition $S(0)=1$. The Hamiltonian, $H$, governing the neutrino flavour evolution through the fluctuating matter is the sum of a constant vacuum term $H_V$ and a term 
coming from the effect of matter $H_{M}$ \cite{Wolfenstein1977,M&S1986}, that is $H = H_V+H_{M}$. The vacuum Hamiltonian in the flavour basis is 
\begin{equation}
H^{(f)}_V = \frac{1}{2E} U_V \left( \begin{array}{*{20}{c}} m_1^2-m_2^2 & 0 & 0 \\ 0 & 0 & 0 \\ 0 & 0 & m_3^2-m_2^2 \end{array} \right) U_V^{\dagger}
\end{equation}
where $U_V$ is the vacuum mixing matrix and $m_i$ the three neutrino masses. We set the squared mass differences $m_1^2-m_2^2 = -7.5 \times 10^{-5}\;{\rm eV^2}$ and $m_3^2-m_2^2 =2.32\times 10^{-3}\;{\rm eV^2}$ which are compatible with the mass-squared differences as given by the Particle Data Group \cite{PDG}. Throughout this paper we choose the neutrino energy $E$ to be $5\;{\rm MeV}$. $U_V$ is parameterized by three mixing angles $\theta_{12}$, $\theta_{13}$ and $\theta_{23}$ - we set all possible phases to zero \cite{2009PhRvD..80e3002K} - and given by 
\begin{eqnarray}
U_V & = & 
\left(\begin{array}{lll} c_{12}c_{13} & s_{12}c_{13} & s_{13}
\\ 
-s_{12}c_{23}-c_{12}s_{13}s_{23}
& c_{12}c_{23}-s_{12}s_{13}s_{23}
& c_{13}s_{23} \\ 
s_{12}s_{23}-c_{12}s_{13}c_{23} &
-c_{12}s_{23}-s_{12}s_{13}c_{23}
& c_{13}c_{23} \end{array}\right) 
\label{eq:U}
\end{eqnarray}
where the notation is that $c_{ij} = \cos\theta_{ij}$ and $s_{ij} = \sin\theta_{ij}$. We take the angles to be $\theta_{12} =34^{\circ}$, $\theta_{13} =9^{\circ}$ and $\theta_{23} =45^{\circ}$ \cite{PDG}. 

\subsection{Two Fourier Modes}

We first consider the case where the matter Hamiltonian $H_M$ is taken to be a constant upon which are superposed two FMs with wavenumbers $q_1$ and $q_2$ not in a rational ratio. The matter is regarded as affecting only the electron flavour type, not the other two flavours. The form of the Hamiltonian in the flavour basis, with the first row/column indicating the electron flavour, is thus 
\begin{equation}
H^{(f)}_M(r) = V_{\star} \left[1 + A_1 \cos(q_1 r + \phi_1) + A_2 \cos(q_2 r + \phi_2)\right] \left( \begin{array}{lll} 1 & 0 & 0 \\ 0 & 0 & 0 \\ 0 & 0 & 0 \end{array} \right)
\end{equation}
with $V_{\star}$ the potential from the constant background, $A_1$ and $A_2$ the amplitudes of the fluctuations. In what follows we set $V_{\star}$ to $V_{\star}=6\times 10^{-25}\;{\rm erg}$ and $\phi_1,\phi_2$ to zero.

\begin{figure}[t]
\centering
\includegraphics[scale=1.2]{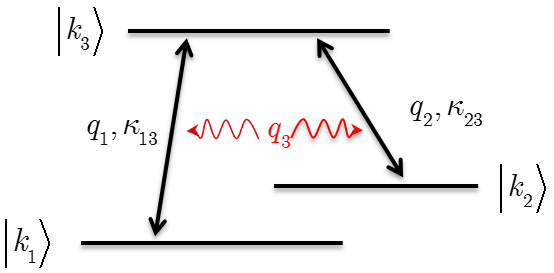}
\caption{The three unperturbed eigenstates and the transitions between them in the two and three FM problems. The modes $q_1$ and $q_2$ are the FMs that drive transitions between the indicates states. The mode $q_3$ is the ``switch mode'' which switches on and off the effect of transitions induced by modes $q_1$ and $q_2$.}
\label{fig:levels}
\end{figure}

The vacuum Hamiltonian and the constant potential $V_{\star}$ form the `unperturbed' Hamiltonian $\breve{H}$. In the flavour basis $\breve{H}$ is not diagonal. We can diagonalize $\breve{H}$ by first finding its matrix of eigenvalues, denoted by $K ={\rm diag}\left( k_1, k_2, k_3\right)$, and then the unitary matrix $\breve{U}$ which satisfies $\breve{H}^{f} = \breve{U}\,K\,{\breve{U}}^{\dagger}$. Since this is standard textbook quantum mechanics, we leave this as an exercise for the reader. For reference, the differences between the eigenvalues are found to be $k_3 - k_1 = 3.835\times 10^{-22}\;{\rm erg}$ and $k_3 - k_2 = 3.715\times 10^{-22}\;{\rm erg}$. Note that since $\breve{H}$ is a function of $V_{\star}$, the eigenvalues and unperturbed mixing matrix, $\breve{U}$ are also functions of $V_{\star}$. The level scheme we end up with is shown in figure (\ref{fig:levels}) with the three eigenstates of the unperturbed system denoted by $\ket{k_1}$, $\ket{k_2}$ and $\ket{k_3}$. 

The two FMs in the matter Hamiltonian are a Fourier-decomposed perturbation
and so we can apply the analytic solution for this kind of perturbation derived in \ref{sec:theory}. The evolution matrix in the basis of the eigenstates of the unperturbed Hamiltonian, denoted as $S^{(\breve{u})}$, is written as the product $S^{(\breve{u})} = \breve{S}\,W\,B$ where $\breve{S}$ is the evolution matrix for the unperturbed states, $W$ is a diagonal matrix designed so as to remove the diagonal elements of the perturbing Hamiltonian in this basis, and $B$ is the evolution matrix which describes the transitions.  
\begin{figure}[b]
\centering
\includegraphics[clip,width=\linewidth]{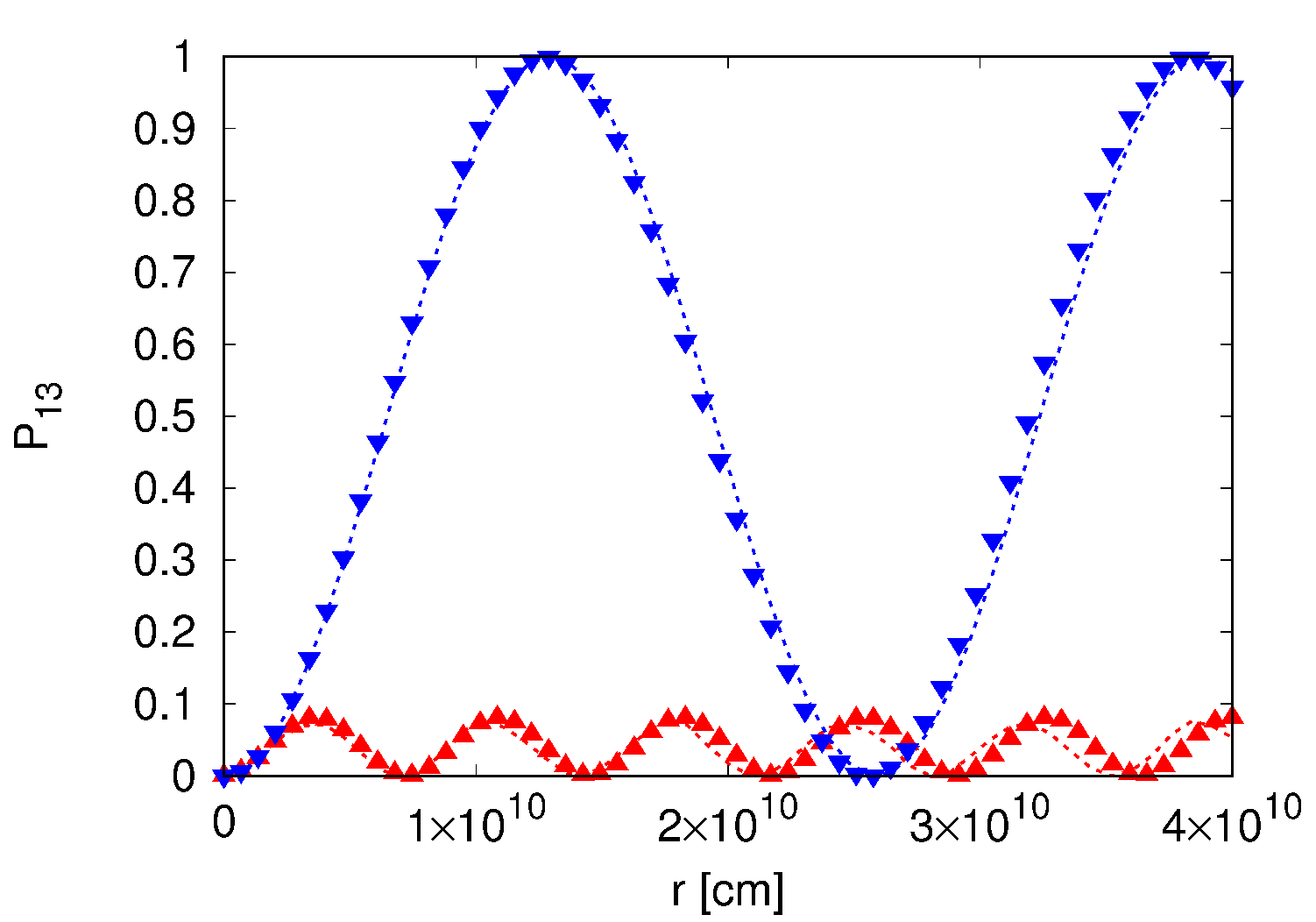}
\caption{The transition probabilities from unperturbed eigenstate 1 to unperturbed eigenstate 3. The blue dashed line is the analytic result for the case $A_1=0.1, A_2=0$ and the red dashed line for $A_1=0.1, A_2=0.5$. The symbols represent the corresponding numerical results.}
\label{fig:ITvsr}
\end{figure}
\begin{figure}[t]
\centering
\includegraphics[clip,width=\linewidth]{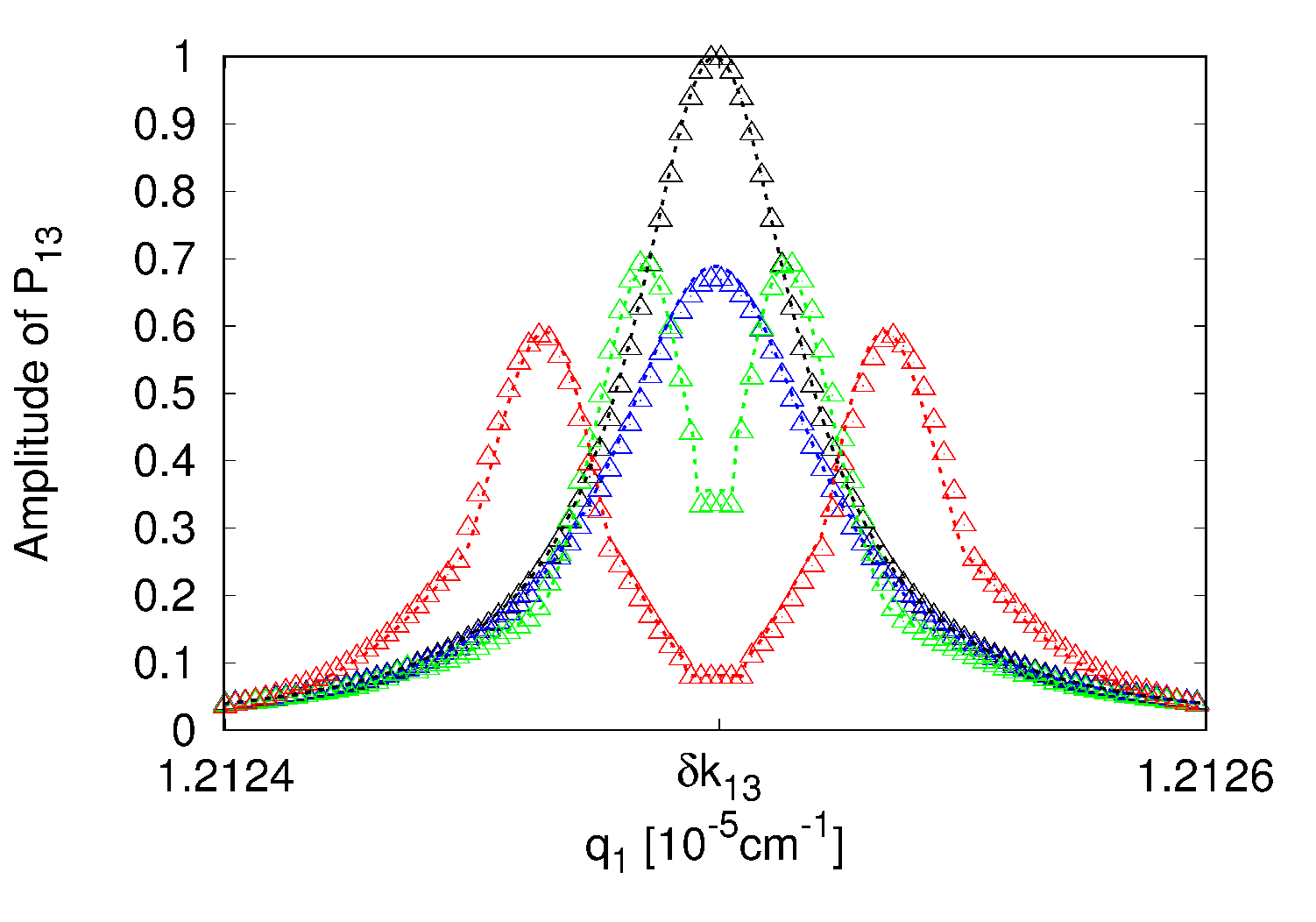}
\caption{The amplitude of $P_{13}$ as a function of $q_1$. The parameters used are $V_{\star}=6\times 10^{-25}\;{\rm erg}, A_1=0.1$, and $A_2=0/0.1/0.2/0.5$ for the black/blue/green/red dashed line and symbols, $A_2=0$ for the blue dashed line and symbols. The symbols represent numerical results, while the dashed lines are from analytic evaluation.}
\label{fig:ITvsq1}
\end{figure}
While this problem may be solved in general, let us consider the case where we set the wavenumbers for the two modes so that $q_1 \approx  k_3 -  k_1$ and $q_2 \approx  k_3 -  k_2$ as shown in figure (\ref{fig:levels}). According to the solution found in \ref{sec:theory}, we must find a set of integers - the RWA integers - for every element $ij$ of the perturbing Hamiltonian for each FM $a$. These integers are labeled $n_{a;ij}$. Given our choice for $q_1$ and $q_2$, the RWA integers we select for the $1,3$ element are $\{ n_{1;13}, n_{2;13} \} = \{ +1 , 0 \}$ and for the $2,3$ element we pick $\{ n_{1;23}, n_{2;23} \} = \{ 0 , +1 \}$. The integer set for the $1,2$ element must therefore be $\{ n_{1;12}, n_{2;12} \} = \{ +1 , -1 \}$ in order that $n_{a;12} + n_{a;23} = n_{a;13}$. The evolution of $B$ is determined by (see equation \ref{dBdt} in appendix)
\begin{equation}
\rmi \frac{\rmd B}{\rmd t} = H^{(B)} B
\end{equation}
where the Hamiltonian $H^{(B)}$ is 
\begin{equation}
\fl
H^{(B)} = \left( \begin{array}{ccc}
    0 & -\rmi \kappa_{12}\, \rme^{\rmi\left(\delta k_{12}+q_1-q_2\right) r} & -\rmi \kappa_{13}\, \rme^{\rmi\left(\delta k_{13} + q_1 \right) r} \\
    \rmi \kappa^{\star}_{12}\, \rme^{-\rmi\left(\delta k_{12}+q_1-q_2\right) r} & 0 & -\rmi \kappa_{23}\, \rme^{\rmi\left(\delta k_{23} + q_2\right) r} \\
    \rmi \kappa^{\star}_{13}\, \rme^{-\rmi\left(\delta k_{13} + q_1\right) r} & \rmi \kappa^{\star}_{23}\, \rme^{-\rmi\left(\delta k_{23} + q_2\right) r} & 0
    \end{array}\right)
\end{equation}
with
\begin{eqnarray}
\kappa_{12} & = & -\frac{2\,\rmi \,G_{1;12}}{z_{1;12}} \,J_{1}\left(z_{1;12}\right) \,J_{1}\left(z_{2;12}\right)
             \;+\; \frac{2\,\rmi \,G_{2;12}}{z_{2;12}} \,J_{1}\left(z_{1;12}\right) \,J_{1}\left(z_{2;12}\right) \label{eq:kappa12:2}\\
\kappa_{13} & = & \frac{2\,\rmi \,G_{1;13}}{z_{1;13}} \,J_{1}\left(z_{1;13}\right) \,J_{0}\left(z_{2;13}\right) \label{eq:kappa13:2}\\
\kappa_{23} & = & \frac{2\,\rmi \,G_{2;23}}{z_{2;23}} \,J_{0}\left(z_{1;23}\right) \,J_{1}\left(z_{2;23}\right) \label{eq:kappa23:2},
\end{eqnarray}
and the quantities $z_{a;ij}$ are defined to be 
\begin{equation}\label{eqn:z}
z_{a;ij} =\frac{ A_a V_{\star} \left( |\breve{U}_{ei}|^2 - |\breve{U}_{ej}|^2 \right) }{2\,q_a} \label{y_a;ij}.
\end{equation}
We notice that both terms in $\kappa_{12}$ are proportional to the product of two Bessel functions $J_1$ so once we recall that the Bessel function $J_{n}(z) \sim z^{|n|}$ for small $z$, we see that the element $\kappa_{12}$ is smaller in magnitude than $\kappa_{13}$ and $\kappa_{23}$ since the values of $z_{a:ij}$ are very small. That is confirmed when we compute the numerical values and find $\kappa_{12} = 6.419\times 10^{-32}\,\rmi\;{\rm erg}$, $\kappa_{13} = -3.888\times 10^{-27}\,\rmi\;{\rm erg}$ and $\kappa_{23} = -1.311\times 10^{-26}\,\rmi\;{\rm erg}$. 

We now proceed to solve for $B$ following the steps found in the appendix.
If we make the approximation that $\kappa_{12}$ is negligibly small compared to $\kappa_{13}$ and $\kappa_{23}$ and that the two wavenumbers are exactly on resonance, $q_1 = k_3 - k_1$, $q_2 = k_3 - k_2$, then we find the analytical expression for the $B$ matrix is \\
\begin{eqnarray}
\fl B = \exp\left(\rmi\,\left[\Lambda-k_3\right]\,r\right)
\left( {\begin{array}{*{20}{c}}
{\frac{{{{\left| {{\kappa _{23}}} \right|}^2}}}{{{Q^2}}} + \frac{{{{\left| {{\kappa _{13}}} \right|}^2}}}{{{Q^2}}}\cos \left( {Qr} \right)}&{\frac{{{\kappa _{13}}\kappa _{23}^*}}{{{Q^2}}}\left[ {\cos \left( {Qr} \right) - 1} \right]}&{ - \frac{{{\kappa _{13}}}}{Q}\sin \left( {Qr} \right)}\\
{\frac{{{\kappa _{23}}\kappa _{13}^*}}{{{Q^2}}}\left[ {\cos \left( {Qr} \right) - 1} \right]}&{\frac{{{{\left| {{\kappa _{13}}} \right|}^2}}}{{{Q^2}}} + \frac{{{{\left| {{\kappa _{23}}} \right|}^2}}}{{{Q^2}}}\cos \left( {Qr} \right)}&{ - \frac{{{\kappa _{23}}}}{Q}\sin \left( {Qr} \right)}\\
{\frac{{\kappa _{13}^*}}{Q}\sin \left( {Qr} \right)}&{\frac{{\kappa _{23}^*}}{Q}\sin \left( {Qr} \right)}&{\cos \left( {Qr} \right)}
\end{array}} \right),\nonumber \\ 
\end{eqnarray}
where $Q^2 = \left|\kappa_{13}\right|^2 + \left|\kappa_{23}\right|^2$. From this result we can extract the transition probability from unperturbed eigenstate 1 to unperturbed eigenstate 3 by taking the squared magnitude of $B_{13}$
\begin{equation}\label{p13}
{P_{13}} = |B_{13}|^2 = \frac{{{{\left| {{\kappa _{13}}} \right|}^2}}}{{{Q^2}}}{\sin ^2}\left( {Qr} \right) = \left( {1 - \frac{{{{\left| {{\kappa _{23}}} \right|}^2}}}{{{Q^2}}}} \right){\sin ^2}\left( {Qr} \right).
\end{equation}
This result is interesting because it indicates the transition probability $P_{13}$ depends upon the wavenumber $q_2$ which is driving transitions from unperturbed eigenstate 2 to unperturbed eigenstate 3. In the extreme case when $\kappa_{23}$ is significantly larger than $\kappa_{13}$, the transition from states 1 to 3 is strongly suppressed. This is an analog of the Electromagnetically Induced Transparency (EIT) - see, for example, \cite{1990PhRvL..64.1107H,1991PhRvL..66.2593B,2003RvMP...75..457L,2005RvMP...77..633F} - in atomic physics where the presence of a second possible transition between atomic levels 2 and level 3 will inhibit the primary transition from atomic level 1 to level 3 leading to little absorption, and thus transparency, for the light frequency corresponding to the energy splitting of level 1 and 3. 

To illustrate this neutrino version of induced transparency, in figure (\ref{fig:ITvsr}) we plot the transition probability as a function of $r$ when the system is at perfect resonance, namely when $q_1 = k_3 - k_1$ and $q_2 = k_3 - k_2$. The reader will observe that indeed, even though the wavenumber $q_1$ is exactly on resonance with the transition between neutrino states 1 and 3, the probability of being in state 3 has a maximum of only $10\%$ when $A_2 \neq 0$. When we remove the second FM $q_2$ the transition probability $P_{13}$ increases to $100\%$. Note also a) that the solution is periodic even though the two wavenumbers $q_1$ and $q_2$ do not form rational ratio, and b) how well the numerical solution to the problem agrees with the analytic solution. The predicted amplitude and the wavenumber match the amplitude and wavenumber of the numerical solution to within a few percent. 

To see the effect of induced transparency more clearly, we fix $q_2$ at the resonance between states 2 and 3 and scan in $q_1$. The solution for $B$ can be found by evaluating the formal solution and from the element $B_{13}$ we extract the transition probability $P_{13}$. In figure (\ref{fig:ITvsq1}) we plot the amplitude of the oscillations in $P_{13}$ as a function of $q_1$. We see that in the presence of mode $q_2$, the transition probability has a peculiar shape with peaks off-resonance and local minimum at the resonance. If we turn off the second perturbing mode by setting $A_2$ to zero we recover the expected shape for a resonance at $q_1$. Again, we find the analytic solution is able to reproduce the shape of $P_{13}$ versus $q_1$ very well at all the values of $A_2$ used. 


\subsection{Three Fourier modes}
\label{opacity}

Now we add a third FM to the perturbing Hamiltonian which we give an amplitude $A_3$ and wavenumber $q_3$. Thus the perturbing Hamiltonian in the flavour basis becomes 
\begin{equation}
\delta {H^{\left( f \right)}}\left( r \right) = V_{\star}\sum\limits_{j = 1}^{3} {{A_j}\cos \left( {{q_j}r + \phi_j} \right)} \left( {\begin{array}{*{20}{c}}
1&0&0\\
0&0&0\\
0&0&0
\end{array}} \right).
\end{equation}
We shall leave $V_{\star}$ unchanged so that the unperturbed Hamiltonian is the same as the previous case of two FMs with the same eigenvalues. 

Let us again set the wavenumbers for the first two modes so that $q_1 = k_3 - k_1$ and $q_2 = k_3 - k_2$ for the level diagram shown in figure (\ref{fig:levels}). Neither $A_1$ nor $A_2$ are zero and $A_2 > A_1$. For a two FM case this choice for the wavenumbers $q_1$ and $q_2$ and ratio of amplitudes would put the system exactly at the midpoint of figure (\ref{fig:ITvsq1}) so the transition probability $P_{13}$ is suppressed even though the wavenumber $q_1$ is exactly on resonance.
We shall not set the mode $q_3$ to a particular value yet but we shall only consider wavenumbers such that $q_3$ is much smaller than $k_3 - k_1$, $k_3 - k_2$ and $k_2 - k_1$. i.e.\ $q_3$ is not on resonance with any pair of eigenvalue splittings. Thus the sets of RWA integers are very similar to the sets for the two FM case: for the $1,3$ element they are $\{ n_{1;13}, n_{2;13}, n_{3;13} \} = \{ +1 , 0, 0 \}$ and for the $2,3$ element we pick $\{ n_{1;23}, n_{2;23}, n_{3;23} \} = \{ 0 , +1, 0\}$. Again, these choices mean the integer set for the $1,2$ element is determined and must therefore be $\{ n_{1;12}, n_{2;12}, n_{3;12} \} = \{ +1 , -1,0 \}$ in order that $n_{a;12} + n_{a;23} = n_{a;13}$. The structure of the Hamiltonian for $H^{(B)}$ is exactly the same as in equation (\ref{eq:hb}). 
The expressions for the $\kappa$'s are:
\begin{eqnarray}
\fl
\kappa_{12} = -\frac{2\,\rmi\,G_{1;12}}{z_{1;12}} \,J_{1}\left(z_{1;12}\right) \,J_{1}\left(z_{2;12}\right) \,J_{0}\left(z_{3;12}\right)
                    \;+\; \frac{2\,\rmi \,G_{2;12}}{z_{2;12}} \,J_{1}\left(z_{1;12}\right) \,J_{1}\left(z_{2;12}\right) \,J_{0}\left(z_{3;12}\right)  \\
\kappa_{13} = \frac{2\,\rmi \,G_{1;13}}{z_{1;13}} \,J_{1}\left(z_{1;13}\right) \,J_{0}\left(z_{2;13}\right) \,J_{0}\left(z_{3;13}\right) \\
\kappa_{23} = \frac{2\,\rmi \,G_{2;23}}{z_{2;23}} \,J_{0}\left(z_{1;23}\right) \,J_{1}\left(z_{2;23}\right) \,J_{0}\left(z_{3;23}\right)
\end{eqnarray}
where $z_{a;ij}$ has the same meaning as for the two FM case, which is defined by equation (\ref{eqn:z}). 
The expressions again show $\kappa_{12}$ is much smaller than $\kappa_{13}$ and $\kappa_{23}$ when $z_{1;ij}$ and $z_{2;ij}$ are small. These expressions look very similar to those given in equations (\ref{eq:kappa12:2}) - (\ref{eq:kappa23:2}) for the two FM case, in fact the only difference is the presence of $J_{0}\left(z_{3;ij}\right)$. But the presence of this new term permits new phenomena because as we vary the wavenumber $q_3$ and/or its amplitude $A_3$ it becomes possible for either $z_{3;13}$ or $z_{3;23}$ to become equal to a zero of the Bessel function $J_0$. The effect will be to either switch off $\kappa_{13}$ or $\kappa_{23}$. If we switch off $\kappa_{13}$ then no transitions between states $1$ and $3$ can occur thus $P_{13}=0$ \emph{even though mode $q_1$ is on resonance}. At this value of $q_3$ the induced transparency effect, which was only partial for the two FM case, will become complete for three FMs. If we switch off $\kappa_{23}$ then the effect of the third FM is to cancel the induced transparency effect and so restore amplitude of the oscillations of $P_{13}$ to 100\%. We call this effect Restored Opacity. In summary, by scanning in the non-resonant mode $q_3$ we can tune the opacity of the system from zero to 100\% even though this mode is nowhere close to being resonant. 
\begin{figure}[t!]
\centering
\includegraphics[clip,width=\linewidth]{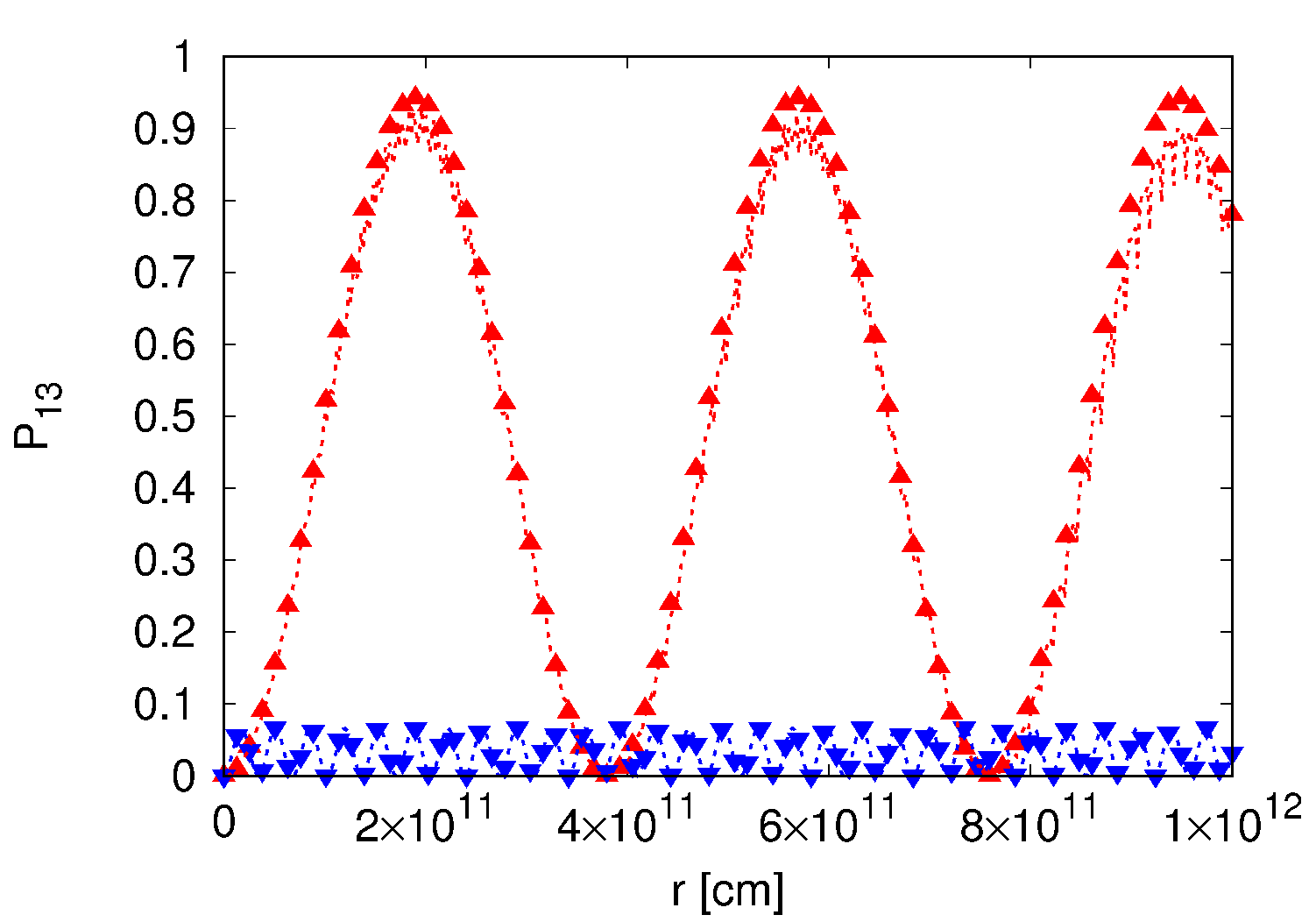}
\caption{The probability $P_{13}$ as a function of $r$ for a three flavour neutrino model. The wavenumbers $q_1$ and $q_2$ are set to $q_1 = k_3 - k_1$ and $q_2 = k_3 - k_2$ with amplitudes $A_1=0.02$ and $A_2=0.1$. The third wavenumber is $q_3=5.24\times10^{-10}{\rm cm^{-1}}$. The dashed lines are the numerical solutions, the triangle symbols are the analytic prediction. The blue curve is for the case $A_3 =0$ and produces an example of Induced Transparency. The red curve is for $A_3=0.2$ and produces an example of Restored Opacity. }
\label{fig:ROvsr}
\end{figure}

To test these predictions we solve the for the transition probability $P_{13}$ numerically making no approximation. As for the two FM case, we set the potential $V_{\star}$ to $V_{\star}=6\times 10^{-25}\;{\rm erg}$ and the wavenumbers $q_1$ and $q_2$ are set to $q_1 = k_3 - k_1$ and $q_2 = k_3 - k_2$ with amplitudes $A_1=0.02$ and $A_2=0.1$. The third wavenumber $q_3$ is set to $q_3=5.24\times10^{-10}{\rm cm^{-1}}$ and we consider two cases: $A_3 =0$ and $A_3 =0.2$. The comparison between the numerical and analytic solutions is shown in figure (\ref{fig:ROvsr}). In the $A_3 =0$ case we expect induced transparency and indeed the figures shows that is correct with very small amplitude oscillations in $P_{13}$ even though the wavenumber $q_1$ is exactly on resonance between those pair of states. When we switch on the third mode we find $z_{3;23}$ is equal to a root of $J_0$ which means $\kappa_{23}=0$. This should return the amplitude of the oscillations of the transition probability $P_{13}$ back to unity and the figure indicates that does indeed occur: the presence of the third FM with this amplitude and wavenumber leads to a restoration of the opacity. 

\begin{figure}[t!]
\centering
\includegraphics[clip,angle=0,width=\linewidth]{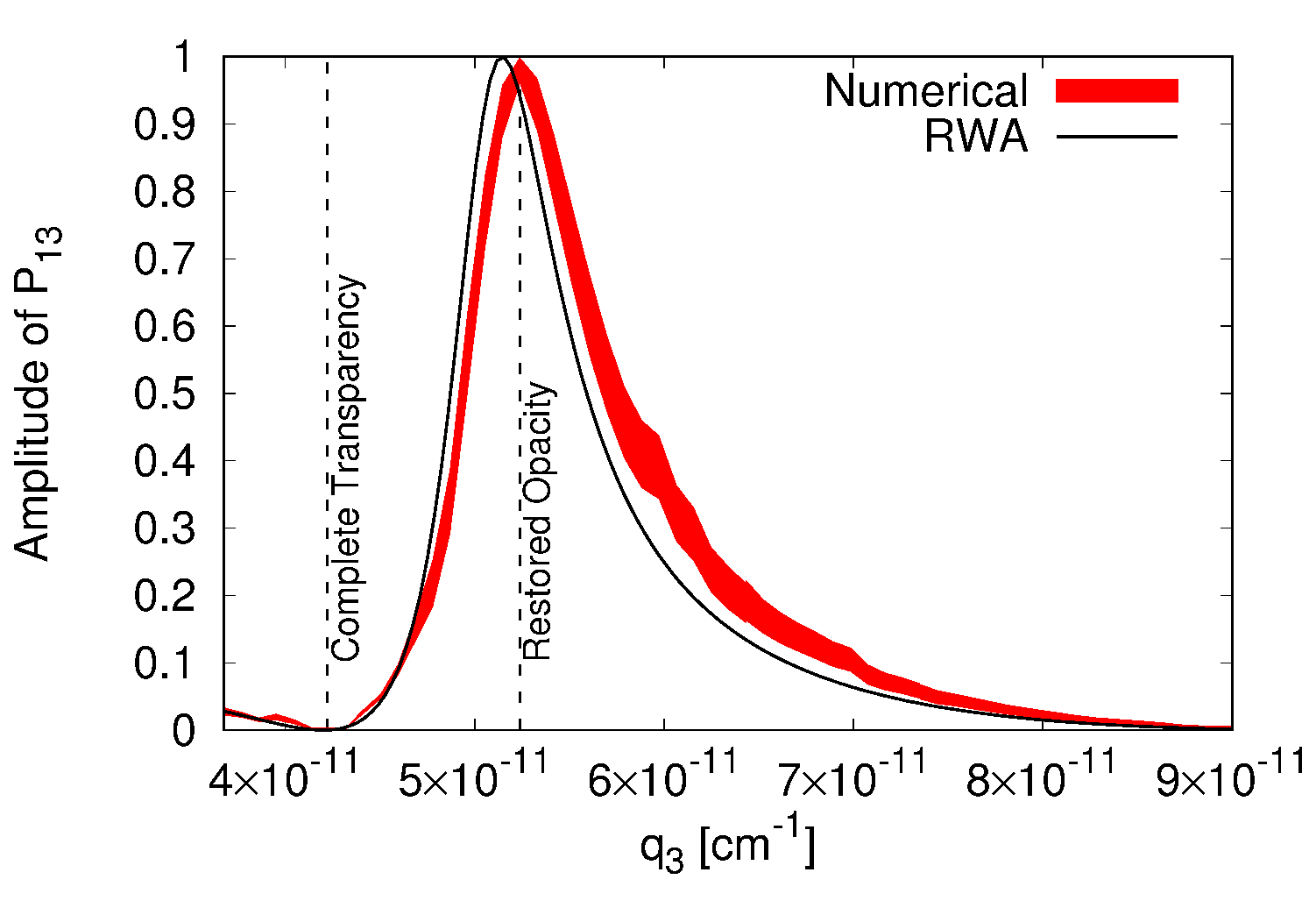}
\caption{The amplitude of $P_{13}$ as a function of the wavenumber $q_3$. The potential $V_{\star}=6\times 10^{-25}\;{\rm erg}$ and the wavenumbers $q_1$ and $q_2$ are set to $q_1 = k_3 - k_1$ and $q_2 = k_3 - k_2$ with amplitudes $A_1=0.002$ and $A_2=0.01$. The amplitude of the third FM $q_3$ is $A_3=0.02$. The black solid line is from analytic evaluation and the thick red line is from the numerical solutions with the thickness of the band indicating the width of the fluctuations, an example of which is shown in figure (\ref{fig:ROvsr}). The values of $q_3$ which give Complete Transparency and Restored Opacity are indicated.}
\label{fig:ROvsq3}
\end{figure}

To further illustrate the power of the third FM, in figure (\ref{fig:ROvsq3}) we fix the amplitudes at $A_1=0.002,A_2=0.01$ and $A_3=0.02$, and scan in the wavenumber $q_3$. The purpose of using smaller amplitudes for the FMs is to suppress the fluctuations of the transition probability seen in the numerical results which make it hard to determine the transition amplitude. Note this choice also makes the corresponding value of $q_3$ which cause the Bessel functions to hit their roots smaller than in the example shown in figure (\ref{fig:ROvsr}). From every numerical solution we fit two sinusoids with amplitudes that enclose the oscillations of $P_{13}$ as seen in figure (\ref{fig:ROvsr}). The spread in amplitudes forms the width of the band for the numerical results shown in figure (\ref{fig:ROvsq3}).
The comparison of the theory and numerical solutions in figure (\ref{fig:ROvsq3}) indicate the theory does a very good job of reproducing the numerical results. At $q_3=5.24\times10^{-11}{\rm cm^{-1}}$, $\kappa_{23}$ is zero and therefore opacity is restored. When $q_3=4.22\times10^{-11}{\rm cm^{-1}}$ we find $z_{3;13}$ is a root of $J_0$ which forces $\kappa_{13}$ to be zero and thus we have complete transparency. 

\section{Turbulence}
\label{turbulence}
We finish by considering the case of neutrino evolution through a turbulent medium. We assume the medium has an average density $\rho_{\star}$ and in what follows we have set the mean density of the medium to be $\rho_{\star} = 100\;\rm{g}/\rm{cm}^3$ which is a typical matter density at $r \sim 10^5\;\rm{km}$ above the proto-neutron star for a supernova in the late cooling phase. 
This density sets the scale $V_{\star}$ for the matter Hamiltonian $H_M$ to be $V_{\star} = \sqrt 2 {G_F} Y_e \rho_{\star} / m_p$ where $G_F$ is the Fermi constant, $m_{p}$ is the proton mass, and $Y_{e}$ is the electron fraction. For our calculations we have adopted $Y_e=0.5$ which is also consistent with the electron fraction found at $r \sim 10^5\;{\rm km}$ in supernova simulations at late times. The matter Hamiltonian $H_M$ in the flavour basis can be written as  
\begin{equation}
{H_M^{\left( f \right)}}\left( r \right) = V_{\star} \left( {1 + \sum\limits_{a = 1}^{{N_q}} {\left\{ {{A_a}\cos \left( {{q_a}r} \right) + {B_a}\sin \left( {{q_a}r} \right)} \right\}} } \right) \left( {\begin{array}{*{20}{c}}
1&0&0\\
0&0&0\\
0&0&0
\end{array}} \right) 
\end{equation}
where the $A_{a}$'s and $B_{a}$'s are the amplitudes of the FMs, the $q_a$'s are the wavenumbers, and $N_{q}$ is the number of FMs. 
We assume the turbulence is a Gaussian random field with a power spectrum $E(q)$ which is an inverse power law i.e.
\begin{equation}
E(q) = \frac{\alpha-1}{2\,q_{\star}}\,\left( \frac{q_{\star}}{q} \right)^{\alpha}\,\Theta(|q|-q_{\star}).
\end{equation}
In this equation $q_{\star}$ is a cut-off scale for the turbulence wavenumbers $q$ and $\alpha$ is the power spectral index. 
The spectral index we use is the Kolmogorov value of $\alpha = 5/3$.
The number of Fourier modes $N_q$ is determined by the dynamic range i.e. the ratio of the largest spatial scale to the smallest. It is found in practice that for every decade of dynamic range, one needs at least 3 wavenumbers in order to reproduce the statistical properties of the field satisfactorily \cite{2007JCoPh.226..897K,2013PhRvD..88b5004K}. 
In our case, we determined $q_{\star}$ and $N_q$ by first finding the eigenvalues $k_1$, $k_2$ and $k_3$ of the unperturbed Hamiltonian. 
We then picked a value for $q_{\star}$ and dynamic range of the turbulence so as to cover the wavenumbers corresponding to the differences of eigenvalues $k_{3}-k_{1}$ and $k_{3}-k_{2}$. We used $q_{\star}=1.0\times 10^{-6}\;\rm{cm}^{-1}$, the dynamic range was 2.5 orders of magnitude and we use $N_q = 40$ FMs. 
An instance of a random field may be constructed by selecting the  amplitudes $A_a$, $B_a$ and the wavenumbers $q_a$ from probability distributions chosen so as to satisfy the chosen statistical properties. For this work the Gaussian random field for the turbulence is generated by the `Algorithm C' from Kramer, Kurbanmuradov and Sabelfeld \cite{2007JCoPh.226..897K} after setting the rms amplitude of the random field to be $0.25$. 

\begin{figure}[t!]
\centering
\includegraphics[width=.48\textwidth]{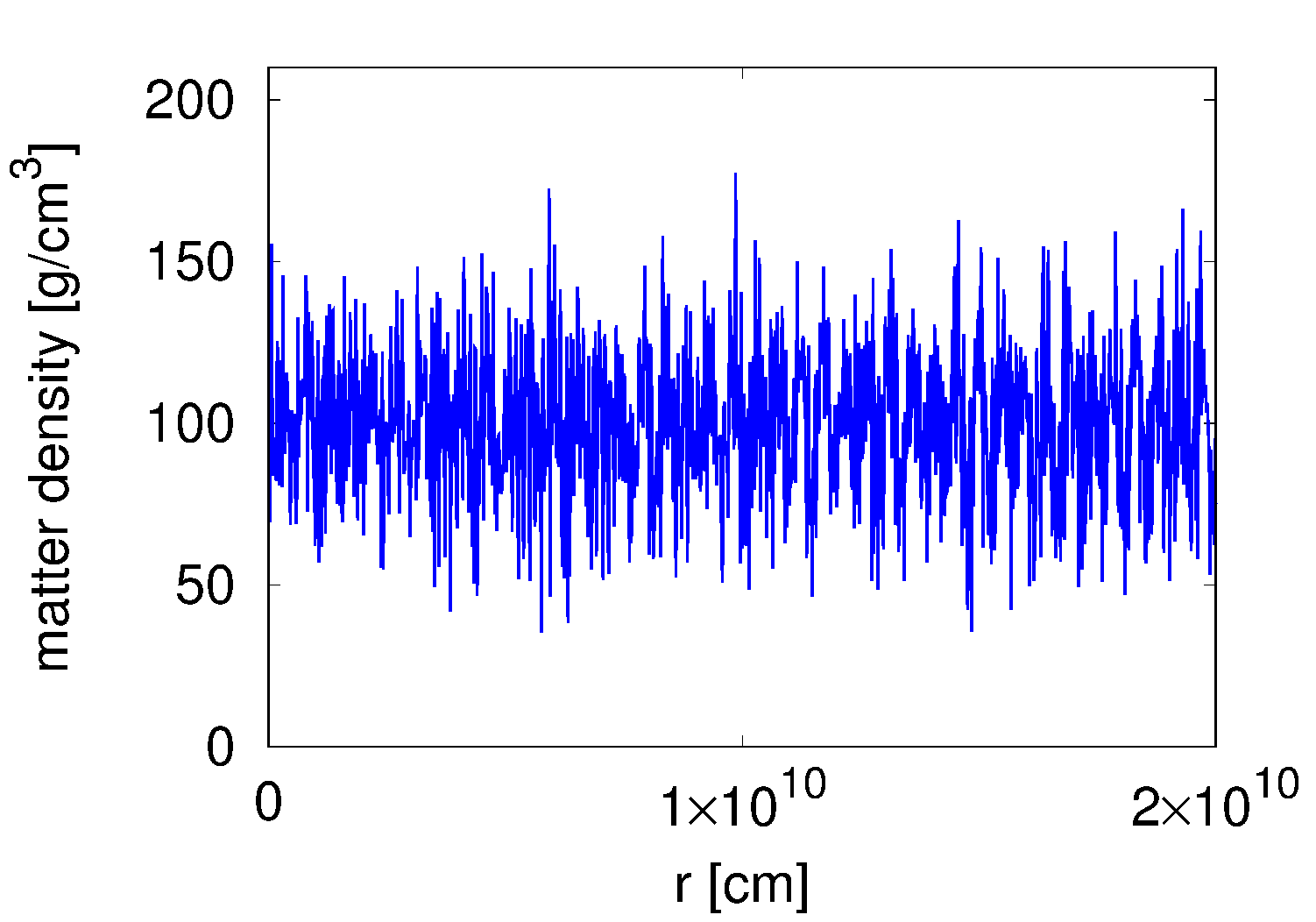}
\includegraphics[width=.48\textwidth]{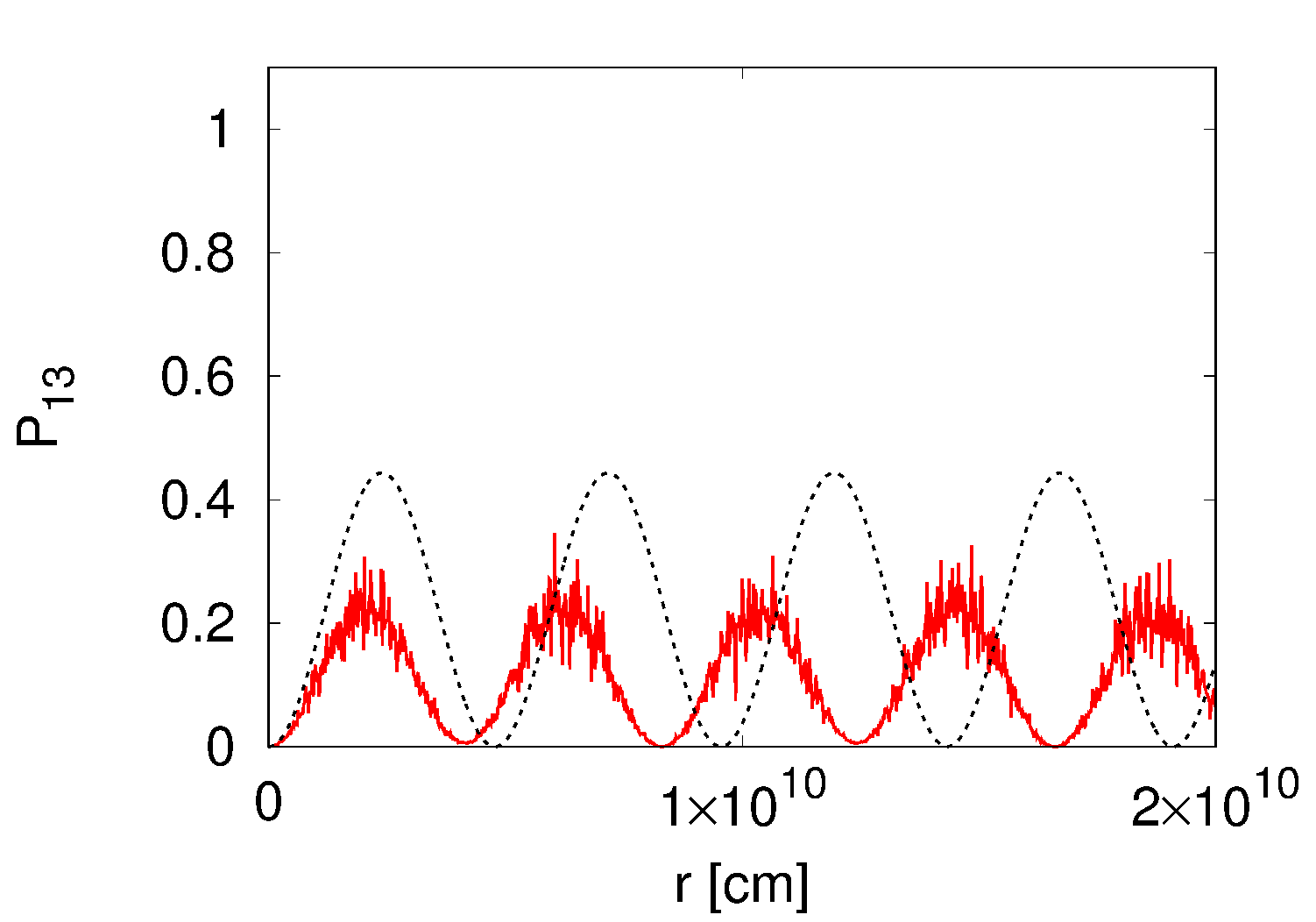}
\caption{Left panel: The turbulence profile with 40 Fourier modes. Right panel: Numerical (red solid) and analytical (black dotted) transition probability with induced transparency.}
\label{fig:turbulence1}
\end{figure}

With all the parameters set, we generated an instance of the turbulence and the wavenumbers and amplitudes of the FMs we obtained are given in \S\ref{FMs}. Before using them we examined the wavenumbers generated by the algorithm and adjusted the two FMs which were closest to the splittings $k_{3}-k_{1}$ and $k_{3}-k_{2}$ in order that these two FMs were exactly equal to the resonant values. The amplitudes were left unchanged. This was done so that we had an instance of turbulence which, according to the two FM case discussed earlier, should be a case of Induced Transparency. The two wavenumbers which were changed are indicated in table (\ref{table:FMs}). With this tweaked instance of turbulence we then proceeded to construct $H_{M}^{(f)}$ and numerically solved the Schr\"{o}dinger equation. The `e-e' element of the matter Hamiltonian and the resulting transition probability $P_{13}$ are shown in figure (\ref{fig:turbulence1}) where we observe that the transition probability $P_{13}$ is suppressed even though the stimulative FM in the turbulence is resonant. Knowing the amplitudes and wavenumbers of the turbulence also allows us to use the analytical theory to make a prediction for this transition probability. The prediction is also shown in figure (\ref{fig:turbulence1}) and we see the theory gives the amplitude qualitatively well and wavenumber to within $\sim 20\%$. 

\begin{figure}[t!]
\centering
\includegraphics[width=.48\textwidth]{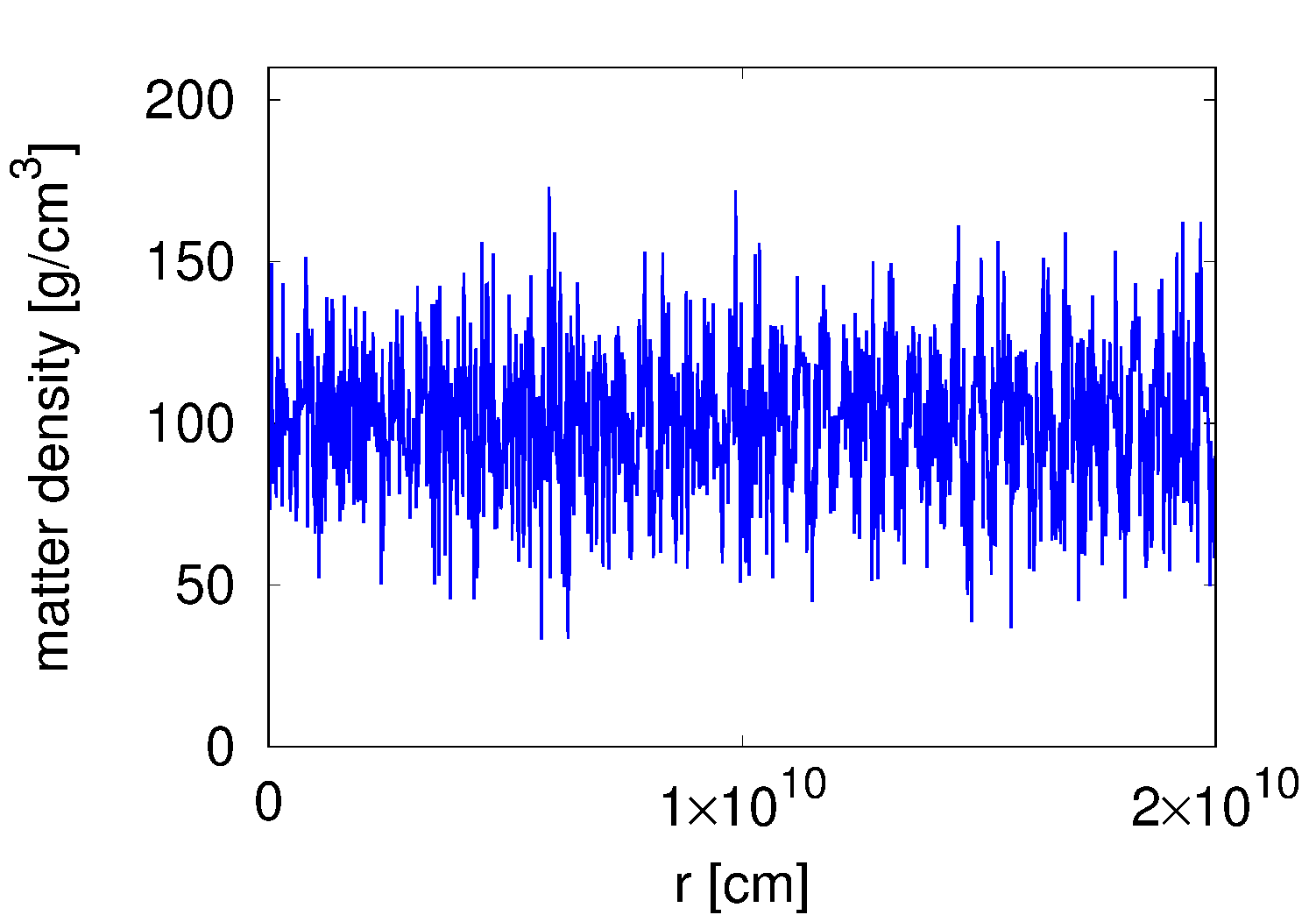}
\includegraphics[width=.48\textwidth]{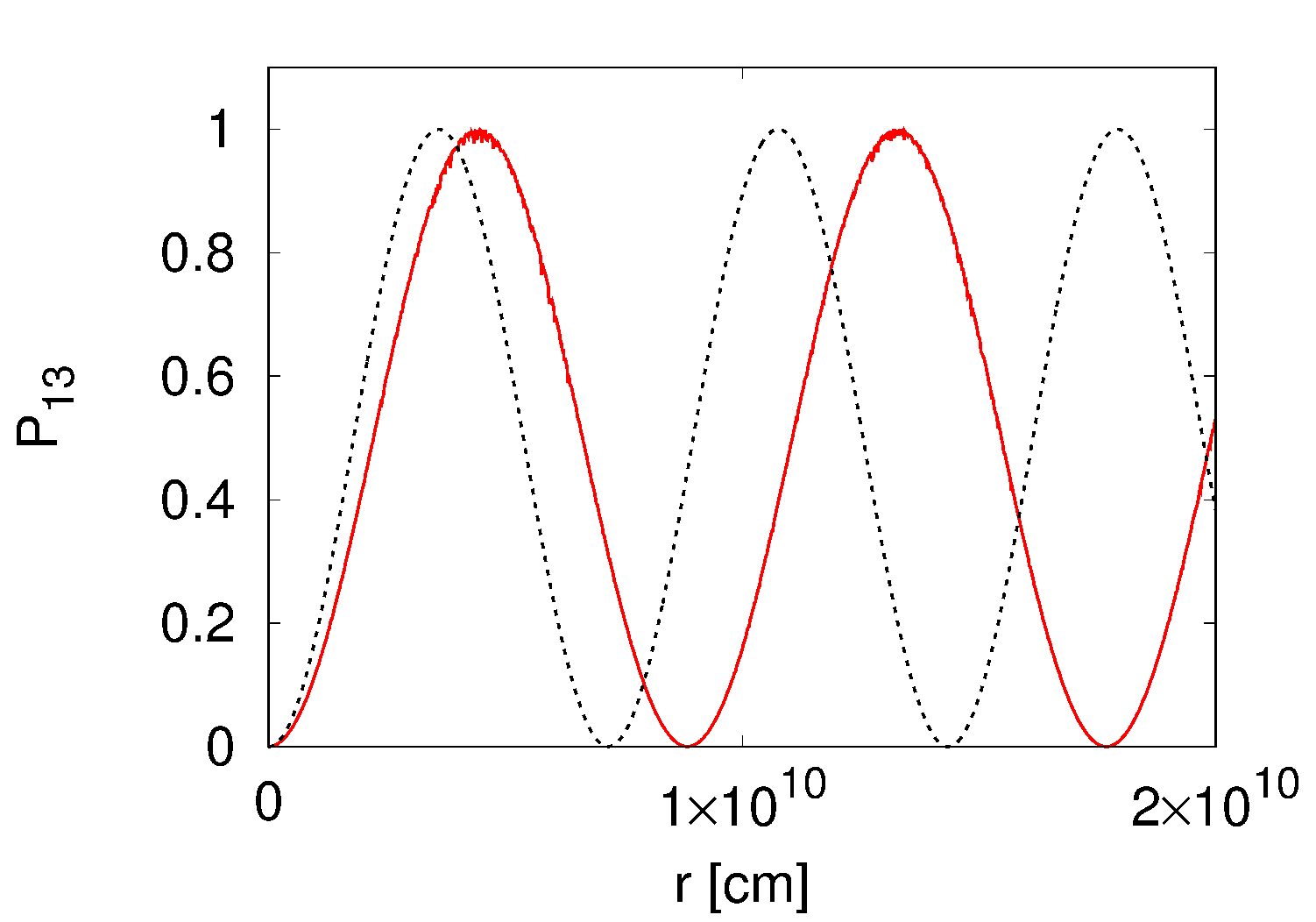}
\caption{Left panel: The same turbulence profile with the suppressive mode removed. Right panel: Numerical (red solid) and analytical (black dotted) transition probability without induced transparency.}
\label{fig:turbulence2}
\end{figure}

In order to verify this was an instance of Induced Transparency, we set to zero the amplitude of the mode which matched the splitting between eigenvalues $k_3 - k_2$. We then reconstructed the matter Hamiltonian $H_M$ and resolved the Schr\"{o}dinger equation. The transition probability is shown in figure (\ref{fig:turbulence2}). Note how the matter Hamiltonian in the left panel is almost identical to that in the left panel of figure (\ref{fig:turbulence1}). Nevertheless, by setting to zero the amplitude of the mode whose wavenumber matches $k_3 - k_2$ we find the amplitude of the oscillations of $P_{13}$ are unity as one would expect for the case of a resonant FM. Again, we are able to use the analytical theory to predict the amplitude and wavenumber of the oscillations of $P_{13}$ and this prediction is also shown in figure (\ref{fig:turbulence2}). The theory predicts the amplitude to be 100\% and the wavenumber is correct to within $\sim 20\%$.

\section{Discussion and Conclusion}
\label{conclusions}

In this paper we have analyzed how a three-flavour neutrino evolves when subject to density fluctuations composed anharmonic FMs. We found effects in these calculations which are not possible with just two neutrino flavours. Using theory based upon the Rotating Wave Approximation, we are able to predict the amplitude and wavenumber of the neutrino transition probabilities between pairs of states to within a few percent. 
As we varied the wavenumbers of the FMs we discovered the equivalent of electromagnetic induced transparency when the expected maximal oscillations between a given pair of neutrino states could be switched off by the presence of a second resonant FM between another pair of states. When we added a third FM we found the neutrino flavour evolution could be further controlled. At one value for the wavenumber the third FM was able to complete the transparency induced by the second but, at another value, we found the opacity could be restored. 

While these cases are non-trivial and produced interesting results, ultimately our goal is to show how more complex cases such as neutrino evolution through turbulence and neutrino self-interaction can be understood in a similar framework. We showed how case of turbulence can be approached by presenting two calculations which were chosen to show the presence of Induced Transparency. We found the analytical theory predicted amplitudes for the transition probabilities through turbulence which were a good match to the numerical solutions and one could therefore understand why the neutrino could respond so differently to what appear to be two, almost identical, instances of turbulence.

The transition to neutrino self-interaction is not as simple because the Hamiltonian is itself a function of the evolution \cite{2006PhRvL..97x1101D,2006PhRvD..74j5010H}. Thus the approach one would need to take is to take the general solution, insert the solution into the actual self-interaction Hamiltonian and demand self-consistency. It is our intention to undertake this application in future work.

\ack
The authors would like to thanks John Thomas, Bob Golub, Laura Clarke, Jason Bochinski and Gail McLaughlin for their helpful comments on this paper. This work was supported at NC State by DOE grants DE-SC0006417 and DE-FG02-10ER41577. \\


\setcounter{section}{1}
\appendix
\section{Stimulated Transformation}
\label{sec:theory}

The problem of the flavour oscillations of a neutrino propagating through a fluctuating medium is a specific case of the more general quantum mechanical problem, namely the time evolution of a multi-level quantum system given a time-dependent perturbation. 
Determining the response of a quantum mechanical system to a time-dependent perturbation is a frequent endeavor of both experiment and theory in a large number of subfields of physics. A number of phenomena have been found to occur in quantum optics, in electronic spin and nuclear magnetic resonance, and in ultracold atoms and molecules to name just a few. Many reviews of driven quantum systems can be found e.g.\ Cohen-Tannoudji \cite{2015PhyS...90h8013C}. There are several analytic approaches to the calculation of the transition probability between the states of the system in textbook literature with the various techniques having strengths and weaknesses depending upon the form of the perturbation. Even when we restrict our attention to harmonic perturbations, one may compute the transition probability (or transition rate) between states using Floquet theory \cite{1965PhRv..138..979S}, the Rotating Wave Approximation (RWA) \cite{1940PhRv...57..522B}, or Fermi's Golden Rule. A comparison between these techniques for a two-level system can be found in Dion \& Hirschfelder \cite{1976acp....35..265D}. 
The approach we adopt to solve this problem is to generalize the method found in  Patton, Kneller \& McLaughlin \cite{2014PhRvD..89g3022P} to an arbitrary number of neutrino flavours and arbitrary, but Fourier decomposed, density fluctuations. While the applications of our solution found in this paper are neutrino related, we recognize the generality of this problem to other fields. Thus, in order to facilitate applications to other quantum systems, the theory is in the most general terms possible without reference to any particular system.

At some initial time $t_1$ we prepare the system in some arbitrary state - represented by a column vector - which we decompose in terms of the $N$ states of some basis $(X)$. The system then evolves to a time $t_2$ at which we decompose the state in terms of the $N$ states of a possibly different basis $(Y)$. The evolution is described by a matrix $S^{(YX)}(t_2,t_1)$ and the transition probabilities are the set of probabilities that the system in a given initial state $x$ of $(X)$ at $t_1$ is detected in the state $y$ of $(Y)$ at $t_2$. These transition probabilities are denoted by $P^{(YX)}_{yx}(t_2,t_1)$ and are related to the elements of $S^{(YX)}$ by $P^{(YX)}_{yx} = |S^{(YX)}_{yx}|^2$. Since $S^{(YX)}$ must be unitary, one needs $N^2$ independent real parameters in order to describe the matrix $S^{(YX)}$ but note only $(N-1)^2$ of the elements of $P^{(YX)}$ are independent. Hereafter we shall work with the case where the bases $(X)$ and $(Y)$ are the same although there are certainly circumstances where knowing the evolution from one basis to a different basis is useful. Note also that throughout this paper we set $\hbar=c=1$.

In the generic basis $(X)$ the evolution matrix can be found by solving the Schr\"{o}dinger equation 
\begin{equation}
\rmi {\frac{\rmd S}{\rmd t}}^{(XX)} = H^{(X)}\,S^{(XX)} \label{eq:dSdt}
\end{equation}
where $H^{(X)}$ is the Hamiltonian in the basis $(X)$. The initial condition is $S^{(XX)}(t_1,t_1) = 1$. We make no assumption about the structure of $H^{(X)}$ except that it be possible to separate the Hamiltonian into an unperturbed piece $\breve{H}^{(X)}(t)$ and a position dependent perturbation $\delta H^{(X)}(t)$ i.e.\ $H^{(X)}(t) = \breve{H}^{(X)}(t) + \delta H^{(X)}(t)$.

If $\breve{H}^{(X)}(t)$ is not diagonal then we introduce an instantaneous unperturbed eigenbasis $(u)$ by finding the unitary matrix $\breve{U}(t)$ defined by $\breve{H}^{(X)} = \breve{U}\,K\breve{U}^{\dagger}$ where $K$ is the diagonal matrix of the eigenvalues of $\breve{H}$, that is $K = \rm{diag}( k_1,k_2,\ldots)$. The evolution matrix in the instantaneous unperturbed eigenbasis is related to the evolution $S^{(XX)}$ by $S^{(uu)}(t_2,t_1) = \breve{U}^{\dagger}(t_2) S^{(XX)}(t_2,t_1)\breve{U}(t_1)$. In this unperturbed eigenbasis
\begin{equation}
H^{(u)} = K - \rmi \breve{U}^{\dagger}\,\frac{\rmd \breve{U}}{\rmd t} + \breve{U}^{\dagger}\delta H^{(X)}\breve{U} 
\end{equation} 
We now write the evolution matrix in the unperturbed eigenbasis as the product $S^{(uu)} = \breve{S}\,A$ where $\breve{S}$ is defined to be the solution of 
\begin{equation}
\rmi \frac{\rmd \breve{S}}{\rmd t} = \left[ K - \rmi \breve{U}^{\dagger}\,\frac{\rmd \breve{U}}{\rmd t} \right] \,\breve{S}.
\end{equation}
If we know the solution to the unperturbed problem, $\breve{S}$, we can solve for the effect of the perturbation by finding the solution to the differential equation for $A$: 
\begin{equation}
\rmi \frac{\rmd A}{\rmd t} = \breve{S}^{\dagger}\,\breve{U}^{\dagger}\delta H^{(X)}\breve{U}\,\breve{S} \,A. \label{dAdt}
\end{equation}
In general the term $\breve{U}^{\dagger}\delta H^{(X)}\breve{U}$ which appears in this equation possesses both diagonal and off-diagonal elements. The diagonal elements are easily removed by writing the matrix $A$ as $A=W\,B$ where $W=\exp(-\rmi\Xi)$ and $\Xi$ a diagonal matrix $\Xi=\rm{diag}(\xi_{1},\xi_{2},\ldots)$. Substitution into (\ref{dAdt}) gives a differential equation for $B$
\begin{equation}\label{dBdt}
\rmi \frac{\rmd B}{\rmd t} = W^{\dagger}\left[\breve{S}^{\dagger}\breve{U}^{\dagger}\delta H^{(X)}\breve{U}\,\breve{S} -\frac{\rmd \,\Xi}{\rmd t}\right]\,W\,B \equiv H^{(B)} B
\end{equation}
and $\Xi$ is chosen so that $d\,\Xi/dt$ removes the diagonal elements of $\breve{S}^{\dagger}\breve{U}^{\dagger}\delta H^{(X)}\breve{U}\,\breve{S}$. Once $\Xi$ has been found, determining transition probabilities is reduced to solving for the $B$ matrix.

%
%

\subsection{Fourier-decomposed Perturbations}

We now consider the specific case of a constant potential for $\breve{H}^{(X)}$. This form for $\breve{H}$ means $\breve{S}$ is a diagonal matrix $\breve{S}=\exp(-\rmi  K\,t)$. The perturbation $\delta H$ is taken to be a Fourier-like series of the form 
\begin{equation}\label{HM} 
 \delta H^{(X)} = \sum_a (C_{a}\,\rme^{\rmi q_{a} t} + C_{a}^{\dagger}\,\rme^{-\rmi q_{a} t})
\end{equation}
where $C_a$ is an arbitrary complex matrix and $q_a$ the frequency of the $a^{th}$ FM of the perturbation. We make no restriction on the number of FMs, the frequencies $q_a$ nor the size or structure of the matrices $C_a$. This generalization to arbitrary structure for the $C_a$'s is where we depart from previous analyses by PKM \cite{2014PhRvD..89g3022P}. We also refer the reader to Brown, Meath \& Tran \cite{Brown:Meath:Tran} and Avetissian, Avchyan \& Mkrtchian \cite{2012JPhB...45b5402A} who considered the related but simpler problem of the effect of two lasers of different colors, i.e.\ two FMs, upon a two-level dipolar molecule.

Given this form for the perturbation, equation (\ref{dBdt}) indicates we need to consider the combination $\breve{U}^{\dagger}C_{a}\breve{U}$. If we write the diagonal elements of $\breve{U}^{\dagger}C_{a}\breve{U}$ as 
\begin{equation}
 {\rm diag}( \breve{U}^{\dagger}C_a\breve{U} ) = \frac{F_a}{2\,\rmi}  \exp(\rmi\,\Phi_a)
 \label{eq7}
\end{equation}
where $F_a$ is a diagonal matrix of amplitudes $F_a={\rm diag}(f_{a;1},f_{a;2},\ldots)$ and $\Phi_a$ the diagonal matrix of phases $\Phi_a={\rm diag}(\phi_{a;1},\phi_{a;2},\ldots)$, then the matrix $\Xi$ is found to be $\Xi(t) = \sum_a \Xi_a(t)$
with 
\begin{equation}
\Xi_a(t) = \frac{F_a}{q_a} \big[ \cos\Phi_a - \cos(\Phi_a + q_a\,t) \big].
\end{equation} 
We denote the diagonal elements of $\Xi_a$ as $\xi_{a;1}, \xi_{a;2},\ldots$. Next we rewrite the off-diagonal elements of $\breve{U}^{\dagger}C_a\breve{U}$ as a matrix $G_a$ i.e\ $G_a ={\rm offdiag}(\breve{U}^{\dagger}C_a\breve{U})$. Putting together the solution for $\Xi$ and $\breve{S}$ and inserting the new matrix $G_a$, we find the Hamiltonian for $B$ is 
\begin{equation}
\fl
 H^{(B)} =\exp(\rmi\,\Xi)\exp(\rmi\,K\,t)\,\Big(\sum_a \left[ G_a \rme^{\rmi q_a\,t} + G^{\dagger}_a \rme^{-\rmi q_a\,t}  \right] \Big)\, \exp(-\rmi\,K\,t) \exp(-\rmi\,\Xi) 
\end{equation} 
Written explicitly the element $ij$ of the Hamiltonian is
\begin{equation}
 H_{ij}^{(B)} = \rme^{\rmi (\delta k_{ij}t+\delta\xi_{ij})} \sum_a\Big[ G_{a;ij}\rme^{\rmi q_a\,t} + G^{\star}_{a;ji}\rme^{-\rmi q_a\,t}\Big]
\end{equation} 
where $\delta k_{ij} = k_{i}-k_{j}$ and $\delta\xi_{ij} = \xi_{i}-\xi_{j}$. 
\\
\\
If we define 
\begin{eqnarray}
x_{a;ij} & = & \frac{f_{a;i}}{q_a}\cos\phi_{a;i} - \frac{f_{a;j}}{q_a}\cos\phi_{a;j} \label{eq11}\\
y_{a;ij} & = & \frac{f_{a;i}}{q_a}\sin\phi_{a;i} - \frac{f_{a;j}}{q_a}\sin\phi_{a;j}  \label{eq12}\\
z_{a;ij} & = & \sqrt{x_{a;ij}^{2}+y_{a;ij}^{2}}   \label{eq13}\\
\psi_{a;ij} & = & \arctan\left( {\frac{{{y_{a;ij}}}}{{{x_{a;ij}}}}} \right) \label{eq14}
\end{eqnarray}
then the term $\delta\xi_{ij}$ is equal to

\begin{equation}
 \delta\xi_{ij} = \sum_a\Big[ x_{a;ij} - z_{a;ij}\cos(q_a\,t+\psi_{a;ij} )\Big].
\end{equation}
The presence of $y_{a;ij}$ in these equations is a new feature of the more general perturbing Hamiltonian we are considering. We now make use of the Jacobi-Anger expansion for $\rme^{\rmi\delta\xi_{ij}}$ 
\begin{equation}
\rme^{\rmi\delta\xi_{ij}} = \prod_a \left\{ \rme^{\rmi x_{a;ij}} \sum_{m_a=-\infty}^{\infty} (-\rmi)^{m_a} J_{m_a}(z_{a;ij})\exp\Big[\rmi\,m_a\left(q_a\,t + \psi_{a;ij}\right)\Big] \right\}.
\end{equation}
If we substitute this expansion into the expression for the elements of $H^{(B)}$ and define $\mu_{a,m_a;ij}$ and $\lambda_{a,m_a;ij}$ to be
\begin{equation}
\fl\lambda_{a,m_a;ij} = (-\rmi)^{m_a}\,\rme^{\rmi x_{a;ij}}\,J_{m_a}(z_{a;ij})\,\rme^{\rmi m_a\psi_{a;ij}}
\label{eq17}
\end{equation}
\begin{equation}
\fl\mu_{a,m_a;ij} 
= (-\rmi)^{m_a}\,\rme^{\rmi x_{a;ij}}\,\left[ G^{\star}_{a;ji}\,J_{m_a+1}(z_{a;ij})\,\rme^{\rmi(m_a+1)\psi_{a;ij}}-G_{a;ij}\,J_{m_a-1}(z_{a;ij})\,\rme^{\rmi(m_a-1)\psi_{a;ij}}\right]
\end{equation}
then we find the element $ij$ of the Hamiltonian is given by 
\begin{equation}\label{HB_ij}
 H_{ij}^{(B)} = \rmi \sum_a\left\{ \sum_{m_a}\mu_{a,m_a;ij}\,\rme^{\rmi ( m_a q_a + \delta k_{ij}) t}\prod_{b\neq a}\left[ \sum_{m_b} \lambda_{b,m_b;ij}\,\rme^{\rmi m_b q_b t} \right]\right\}. 
\end{equation} 


\subsection{Rotating Wave Approximation}

Even though we started with a very general perturbing Hamiltonian, we have found a form for $H^{(B)}$ which has the same structure as that found by PKM. From here on, we follow the same procedure to solve for the matrix $B$. First we adopt the Rotating Wave Approximation. The RWA amounts to selecting a particular value for the integers $m_a$ and $m_b$ in equation (\ref{HB_ij}) and dropping all others.
We do not specify a procedure for selecting those integers though algorithms exist. We expect there is not one procedure that can be adopted universally for all situations. There are some restrictions to be placed on the selection of the integers. In order that the resulting Hamiltonian be solvable we cannot make choices for $m_a$ and $m_b$ for every element $ij$ independently. Only $N-1$ elements are to be regarded as independent and a suitable set could be either those on the sub/superdiagonal or the off-diagonal elements in a particular row or column. The values of  $m_a$ and $m_b$ we select will be different for each independent element. We denote these integers by $n_{a;ij}$ since they are specific both to the frequency $a$ and the element of the Hamiltonian $ij$, and define 
\begin{equation}
 \kappa_{ij} = \sum_a\mu_{a,n_{a;ij};ij} \prod_{b\neq a} \lambda_{b,n_{b;ij};ij} 
\end{equation}
then $H_{ij}^{(B)}$ is simplified to
\begin{equation}
H_{ij}^{(B)} = -\rmi \kappa_{ij} \exp\left[\rmi \left( \sum_a n_{a;ij}\, q_a + \delta k_{ij}\right) t\right],
\end{equation} 
or, in full matrix form 
\begin{eqnarray}
\fl
H^{(B)} = \left( \begin{array}{cccc}
0 & - \rmi \kappa_{12}\,\rme^{\rmi \left[ \delta k_{12} + \sum\limits_a n_{a;12}\,q_a \right]t} & - \rmi \kappa_{13}\,\rme^{\rmi \left[ \delta k_{13} + \sum\limits_a n_{a;13}\,q_a \right]t} & \ldots \\
\rmi \kappa^{\star}_{12}\,\rme^{-\rmi \left[ \delta k_{12} + \sum\limits_a n_{a;12}\,q_a \right]t} & 0 & - \rmi \kappa_{23}\,\rme^{\rmi \left[ \delta k_{23} + \sum\limits_a n_{a;23}\,q_a \right]t} & \ldots \\
\rmi \kappa^{\star}_{13}\,\rme^{-\rmi \left[ \delta k_{13} + \sum\limits_a n_{a;13}\,q_a \right]t} & \rmi \kappa^{\star}_{23}\,\rme^{-\rmi \left[ \delta k_{23} + \sum\limits_a n_{a;23}\,q_a \right]t} & 0 & \ldots \\
 \vdots & \vdots & \vdots & \ddots 
\end{array} \right) \nonumber \\
\label{eq:hb}
\end{eqnarray}
Again, we remind the reader that, for example, only $n_{a;12}$, $n_{a;23}$, $n_{a;34}$ etc.\ are independent: in all other cases the integer $n_{a;ij} = n_{a;i\ell} + n_{a;\ell j}$. As shown by PKM, with this simplified Hamiltonian, equation (\ref{dBdt}), can be solved for the evolution matrix $B$ and we reproduce their solution here for completeness. Since both $n_{a;ij} = n_{a;i\ell} + n_{a;\ell j}$ and $\delta k_{ij} = \delta k_{i\ell} + \delta k_{\ell j}$, we can factorize $H^{(B)}(t)$ into the form $H^{(B)}(t) = \Upsilon(t)\,M\,\Upsilon^{\dagger}(t)$ where the matrix $M$ is a constant, i.e. it contains the couplings $\kappa_{ij}$ only. The matrix $\Upsilon$ is of the form $\Upsilon(t)=\exp(\rmi\,\Lambda\,t)$, where $\Lambda$ is also a constant matrix that depends only on $\delta k_{ij}$, the integer sets $\{n_{a;ij}\}$ and the frequencies $q_a$. Explicitly we can write 
\begin{equation}\label{eq for M}
M = \left( \begin{array}{cccc}
    0 & -\rmi \kappa_{12} & -\rmi \kappa_{13} & \ldots \\
    \rmi \kappa^{\star}_{12} & 0 & -\rmi \kappa_{23} & \ldots \\
    \rmi \kappa^{\star}_{13} & \rmi \kappa^{\star}_{23} & 0 & \ldots \\
    \vdots & \vdots & \vdots & \ddots
    \end{array}\right),
\end{equation}
and one possible choice for the matrix $\Lambda$ is 
\begin{equation}
\Lambda  = \left( {\begin{array}{cccc}
 k_{1} + \sum\limits_a n_{a;1}\,q_a & 0 & 0 & \ldots \\
0 &  k_{2} + \sum\limits_a n_{a;2}\,q_a & 0 & \ldots \\
0 & 0 &  k_{3} + \sum\limits_a n_{a;3}\,q_a & \ldots \\
 \vdots & \vdots & \vdots & \ddots 
\end{array}} \right),
\end{equation}
where $n_{a;i}$ are integers chosen so that $n_{a;i} - n_{a;j} = n_{a;ij}$. Using this factorization of $H^{(B)}(t)$ we find equation (\ref{dBdt}) can be rewritten as   
\begin{equation}
\rmi\Upsilon^{\dagger}\frac{\rmd B}{\rmd t} =  M\Upsilon^{\dagger}\,B
\end{equation}
Instead of solving for $B$ we solve for the combination $\Omega=\Upsilon^{\dagger} B$. The differential equation for $\Omega$ is found to be 
\begin{equation}
\rmi\frac{\rmd \Omega}{\rmd t} = \left(M +\Lambda \right)\,\Omega = H^{(\Omega)}\,\Omega.
\end{equation}
Since the both $M$ and $\Lambda$ are constant matrices, the matrix $H^{(\Omega)}$ is also independent of $t$ meaning $\Omega$ has the formal solution $\Omega(t) = \exp(-\rmi H^{(\Omega)} t)\,\Omega(0)$. The solution for $B$ is thus
\begin{equation}
B(t) = \Upsilon(t)\,\exp(-\rmi H^{(\Omega)} t)\,\Upsilon^{\dagger}(0) B(0). \label{eq:soln for B}
\end{equation}
Now that we have the solution for $B$, the full evolution matrix in the basis $(u)$ is $S=\breve{S}\,W\,B$ but given that both $\breve{S}$ and $W$ are diagonal matrices, the transition probability between certain unperturbed eigenstates is simply the square magnitude of the corresponding off-diagonal element of $B$. \\


\section{Fourier modes in turbulence}
\label{FMs}

In this appendix we list in table (\ref{table:FMs}) the wavenumbers and amplitudes of the 40 Fourier modes used to generate the turbulence in section \ref{turbulence}. To plot figure \ref{fig:turbulence1} the 2 boxed wavenumbers in the list are replaced by the resonant frequencies $q_{2}=k_{3}-k_{2}=1.1681\times 10^{-5}$ and $q_{1}=k_{3}-k_{1}=1.2036 \times 10^{-5}$ respectively, with their amplitudes unchanged. To plot figure \ref{fig:turbulence2} we further set the amplitudes of mode $q_{2}$ to zero.
\begin{table}[h]
\centering
\begin{tabular}
{|P{3cm}|P{3cm}|P{3cm}|}
\hline
$q_{a} \;(\rm{cm}^{-1})$ & $A_{a}$ & $B_{a}$ \\
\hline
$1.1548\times 10^{-06}$ & 0.084032739  & 0.110797904  \\
$1.2470\times 10^{-06}$ & 0.036748031  & -0.05279767  \\
$1.4191\times 10^{-06}$ & 0.026151269  & -0.00429985  \\
$1.6483\times 10^{-06}$ & -0.033460709 & -0.064004617 \\
$2.0299\times 10^{-06}$ & -0.010358338 & 0.064949167  \\
$2.1468\times 10^{-06}$ & -0.018580739 & -0.034739205 \\
$2.6630\times 10^{-06}$ & 0.07392128   & 0.060105807  \\
$2.9946\times 10^{-06}$ & -0.049428249 & -0.049460748 \\
$3.3716\times 10^{-06}$ & -0.026083902 & -0.009837658 \\
$3.7705\times 10^{-06}$ & 0.011933982  & 0.008570219  \\
$4.4115\times 10^{-06}$ & -0.064510778 & -0.130553243 \\
$5.1597\times 10^{-06}$ & -0.017154243 & -0.003635296 \\
$6.1799\times 10^{-06}$ & 0.016785978  & -0.016369856 \\
$7.4943\times 10^{-06}$ & -0.050955539 & 0.01868823   \\
$7.7254\times 10^{-06}$ & 0.018607067  & -0.045718571 \\
$8.9927\times 10^{-06}$ & 0.026290072  & 0.005495659  \\
\framebox{$1.0810\times 10^{-05}$} & -0.057045473 & -0.01573937  \\
\framebox{$1.1872\times 10^{-05}$} & 0.034898257  & -0.007159635 \\
$1.4585\times 10^{-05}$ & 0.006395625  & 0.084931522  \\
$1.7700\times 10^{-05}$ & -0.01469274  & 0.041140348  \\
$1.8033\times 10^{-05}$ & 0.004427475  & 0.005041208  \\
$2.0788\times 10^{-05}$ & 0.057530192  & -0.022535312 \\
$2.7013\times 10^{-05}$ & -0.007757007 & 0.008515802  \\
$3.0258\times 10^{-05}$ & 0.046753954  & 0.025154408  \\
$3.4924\times 10^{-05}$ & -0.081259646 & -0.00970601  \\
$4.1797\times 10^{-05}$ & 0.023584644  & 0.046950117  \\
$4.4670\times 10^{-05}$ & -0.003092699 & -0.026764968 \\
$5.1604\times 10^{-05}$ & 0.000995038  & 0.011487996  \\
$6.3819\times 10^{-05}$ & -0.011470391 & -0.023010379 \\
$6.7349\times 10^{-05}$ & -0.015494087 & -0.02627463  \\
$7.9185\times 10^{-05}$ & 0.00617285   & -0.016379354 \\
$9.9328\times 10^{-05}$ & -0.043149466 & -0.007963938 \\
$1.1079\times 10^{-04}$ & 0.011375218  & -0.025264513 \\
$1.2181\times 10^{-04}$ & -0.005364102 & 0.014622629  \\
$1.5345\times 10^{-04}$ & -0.002306555 & -0.032303389 \\
$1.6665\times 10^{-04}$ & 0.006879804  & 0.009061726  \\
$2.0381\times 10^{-04}$ & 0.018829845  & -0.010870548 \\
$2.1550\times 10^{-04}$ & 0.02317589   & 0.001226525  \\
$2.6244\times 10^{-04}$ & 0.010757126  & -0.003705208 \\
$3.1553\times 10^{-04}$ & -0.005535837 & 0.001056426 \\
\hline
\end{tabular}
\caption{The 40 Fourier modes used to generate the turbulence.}
\label{table:FMs}
\end{table}


\end{document}